\documentclass[aps, prd, 10pt, twocolumn, superscriptaddress,noshowpacs, preprintnumbers, longbibliography, nofootinbib,bibnotes,hyperref,floatfix]{revtex4-2}

\newcommand{\env}[1]{\texttt{#1}}
\newcommand{\PS}[1]{{\color{blue} #1}}

\usepackage{amsmath}
\usepackage{mathtools}
\usepackage{graphicx}
\usepackage{longtable}
\usepackage{bm}
\usepackage{amsfonts}
\usepackage{subfigure}
\usepackage{multirow}
\usepackage{float}
\usepackage{wrapfig}
\usepackage{array}
\usepackage{multimedia}
\usepackage{xcolor}
\usepackage[normalem]{ulem}

\begin{document}

 \title{Gamma-rays and neutrinos from  supernovae of Type Ib/c with late time emission}

\author{Prantik Sarmah}
\email{prantik@iitg.ac.in}
\affiliation{Indian Institute of Technology Guwahati,
Guwahati, Assam-781039, India}

\author{Sovan Chakraborty}
\email{sovan@iitg.ac.in}
\affiliation{Indian Institute of Technology Guwahati,
Guwahati, Assam-781039, India}

\author{Irene Tamborra}
\email{tamborra@nbi.ku.dk}
\affiliation{Niels Bohr International Academy and DARK, Niels Bohr Institute, University of Copenhagen Blegdamsvej 17, 2100, Copenhagen, Denmark}

\author{Katie Auchettl}
\email{katie.auchettl@unimelb.edu.au}
\affiliation{School of Physics, The University of Melbourne, Parkville, VIC 3010, Australia}
\affiliation{ARC Centre of Excellence for All Sky Astrophysics in 3 Dimensions (ASTRO 3D)}
\affiliation{Department of Astronomy and Astrophysics, University of California Santa Cruz, CA
95064, USA}

\date{\today}%

\begin{abstract}
   Observations of some supernovae (SNe), such as SN 2014C, in the X-ray and radio wavebands revealed a rebrightening over a timescale of about a  year since their detection. Such a discovery hints  towards the  evolution of a hydrogen-poor  SN of Type Ib/c into a hydrogen-rich  SN of Type IIn, the late time activity being attributed to the interaction of the SN ejecta with a dense hydrogen-rich circumstellar medium (CSM) far away from the stellar core.   We compute the neutrino and gamma-ray emission from these SNe, considering interactions between the shock accelerated protons and the non-relativistic CSM protons. Assuming  three  CSM models inspired by recent electromagnetic observations,  we explore the dependence of the expected multi-messenger signals on the CSM characteristics. The detection prospects of  existing and upcoming gamma-ray (Fermi-LAT and Cerenkov Telescope Array) and neutrino (IceCube and IceCube-Gen2) telescopes are also outlines. 
    Our findings are in agreement with the non-detection of neutrinos and gamma-rays from past SNe exhibiting late time emission. Nevertheless, the detection prospects of SNe with late time emission in gamma-rays and neutrinos with the Cerenkov Telescope Array and IceCube-Gen2 (Fermi-LAT and IceCube) are promising and could potentially provide new insight into the CSM properties, if the SN burst should occur within $10$~Mpc ($4$~Mpc).

\end{abstract}

\maketitle

\section{Introduction} \label{sec:intro}

Supernovae (SNe)  Ib/c  are among the dominant  SN types ($26\%$)   in the local universe \cite{2011MNRAS.412.1522S}. Typically, the light curve of a  SN Ib/c  fades after a few weeks \cite{2021ApJ...913..145W,2008ApJ...674L..85I,2007ApJ...657L.105F}. However, recent observations of   SN 2014C, a SN of Type Ib/c, have revealed that a fraction of SNe of Type Ib/c exhibits  evidence of  late rebrightening at a few $\mathcal{O}(100)$~days \citep{Milisavljevic:2015bli,Margutti:2016wyh,Tinyanont:2019qol}. Such rebrightening  resembles the behavior  of a hydrogen-rich SN (i.e.~a SN of Type IIn).  Due to this peculiar feature, SN 2014C has been referred to as ``chameleon SN'' \cite{Margutti:2016wyh}.  

The late time (LT) rebrightening may result from the interaction of  the SN ejecta with a dense circumstellar medium (CSM) surrounding the dying star. Observations of SN 2014C suggest that the shock may have  interacted with a dense hydrogen (H) rich CSM located at  larger radii.  
Such CSM structure  could be due to the ejection of the H envelope a few centuries prior to explosion or the interaction of a Wolf-Rayet  star wind with a dense red-supergiant wind~\citep{Brethauer:2020bmy,Margutti:2016wyh}. In addition, evidence for an asymmetric CSM  hints towards an explosion occurring within a binary system \citep[][]{2022ApJ...930...57T,Zapartas:2017zsb}. 
 The dense hydrogen rich CSM of SN2014C is found to be located  at about $10^{16}$--$10^{17}$~cm,  which is at a distance  far away from the stellar envelope ($\sim 10^{11}$~cm) and has a mass of about $1$--$2\ \rm M_{\odot}$ \citep{Margutti:2016wyh,2022ApJ...930..150V,Brethauer:2020bmy}. Such dense CSM has also been observed for different types of  core collapse SNe \citep{Smith:2014txa,Ofek:2013afa,Ofek:2013jua,Margutti:2013pfa}.

For a wind-like CSM, the CSM density depends on the mass-loss rate  ($\dot{M}_{\rm W}$) and the  wind velocity ($v_{\rm W}$). The  CSM of conventional SNe Ib/c in the first few $100$~days (early phase) exhibits  $\dot{M}_{\rm W} \in [10^{-7}, 10^{-4}]~\rm M_{\odot}~yr^{-1}$, with $v_{\rm W}$ being approximately  $10^{2}$--$10^{3}~\rm{km\ s}^{-1}$  \citep{Smith:2014txa}. However, $\dot{M}_{\rm W}$ estimated for SN 2014C  after about $200$--$300$~days (late phase) is  $\mathcal{O}(1)$ $\rm M_{\odot}yr^{-1}$ \citep{Margutti:2016wyh}, with $v_{\rm W} \in [10, 10^3]~\rm{km\ s}^{-1}$ that corresponds to a CSM density $\sim 2 \times 10^6~{\rm cm^{-3}}$ at  $6\times10^{16}~{\rm cm}$ and then falls as a function of the radius as $r^{-2}$.  
Analysis   of the available  X-ray data suggest a constant CSM density up to  $8\times 10^{16}~{\rm cm}$ which then falls following $r^{-2.5}$~\citep{Brethauer:2020bmy}. 
Recent work focusing on  X-ray data from SN 2014C instead infers two different density profiles~\citep[][]{2022arXiv220600842B}. 
One of these density profiles scales as $r^{-1.5\pm 0.01}$, while the other one has a steeper profile falling like $r^{-2.42 \pm 0.17}$. Interestingly, this analysis  reports that the  LT emission from SN 2014C is due to a dense H-rich disk resulting in an asymmetric CSM. These conclusions  are in contrast with the model based on a spherically symmetric CSM density profile falling as $r^{-3}$ \citep[][]{2022ApJ...930..150V}. 
Nevertheless, it is clear that the CSM of SN 2014C is different from the ones usually observed with wind-like CSM (i.e., $r^{-2}$ profile).

Similar LT features have been observed for  SNe 2003gk,  2004cc, 2004dk, 2004gk, and 2019yvr~\citep{Margutti:2016wyh, 2022arXiv220600842B,Mauerhan:2018wes,Kilpatrick:2021hup,2021ApJ...923...32B}; additional examples of past SNe Ib/c showing indirect evidence of similar LT activity have been reported in Refs.~\cite{Margutti:2016wyh,2022arXiv220600842B}. 
All these SNe initially showed properties of usual SNe Ib/c, but later evolved into IIn-like SNe with dense CSM. By relying on current observations, the fraction of  SNe Ib/c with LT emission is expected to be about $2.6\%$ of all core-collapse SNe \citep[][]{Margutti:2016wyh,2014ARA&A..52..487S}.  In the following,   we assume   SN 2014C as representative of this class of chameleon SNe.

The interaction of the SN ejecta  with the CSM may lead to the production of  secondary particles, such as neutrinos and gamma-rays, via   inelastic proton-proton ($pp$) collisions \cite{Kelner:2006tc,2018MNRAS.479.4470M,2014NuPhS.256...94M,2022MNRAS.511.3321C,2007ApJ...657L.105F,2019ApJ...872..157W,2011PhRvD..84d3003M,2014MNRAS.440.2528M,2013PhRvD..88l1301M,Sarmah:2023pld,Kheirandish:2022eox}. The  flux of neutrinos and gamma-rays from the conventional early phase of SNe of Type Ib/c  was found to be  faint, 
 with  poor detection prospects
(the detection horizon being  estimated  to be around $2$--$6$~Mpc)~\citep{Murase:2017pfe,2022JCAP...08..011S,phdthesis,Murase:2018okz}.   However, due to the presence of the dense hydrogen rich CSM at large radii, the fluxes of neutrinos and gamma-rays from SNe Ib/c LT can be  larger than the ones expected in the early SN phase.

Different CSM density profiles may  yield different fluxes of neutrinos and gamma-rays. Therefore, the detection of these secondary particles could be crucial to disentangle the properties of the CSM as well as  probe the shock acceleration mechanism.  In this work, we consider the aforementioned CSM profiles to compute the expected fluxes of neutrinos and gamma-rays  and discuss their detection prospects with current and upcoming gamma-ray (Fermi-LAT and CTA) and neutrino (IceCube and IceCube-Gen2)  telescopes.

This paper is organized as follows. We introduce the  CSM models of the LT emission in Sec.~\ref{sec:CSM}, followed by the modelling of the neutrino and gamma-ray  signals   in Sec.~\ref{sec:model}. The temporal evolution, the spectral energy distribution of the secondaries, and their dependence on the model parameters are explored in Sec.~\ref{sec:tem evolution}. The  detection prospects of SN 2014C-like bursts with current and future gamma-ray and neutrino telescopes are presented in  Sec.~\ref{sec:detection prospects}. Finally, we summarize our findings  in Sec.~\ref{sec:conclusion}.  
{\color{black}The characteristic timescales for proton acceleration and cooling processes are provided in Appendix~\ref{sec:appendix}.}

 \section{Modeling of the circumstellar medium}\label{sec:CSM}
Our understanding of the CSM density profile of SN 2014C is still uncertain and different scenarios have been proposed in the literature \citep[][]{Margutti:2016wyh,Brethauer:2020bmy,2022ApJ...930..150V,2022arXiv220600842B}. 
In this paper, we consider the following   CSM models: 

\begin{itemize}
    \item Model A---A spherically symmetric and dense CSM. The CSM density is assumed to be constant ($n_{\rm CSM} \simeq 10^6~\rm cm^{-3}$) between the inner radius, $r_{\rm i}$ $\simeq 6\times10^{16}$~cm and the  break radius $r_{\rm b}$ $\simeq 8 \times 10^{16}$ cm~\citep[][]{Brethauer:2020bmy}. The CSM density beyond $r_{\rm b}$ falls  as $r^{-2.5}$ up to the outer radius, $r_{\rm o}$ $\simeq 2.5 \times 10^{17}$ cm. 
The origin of the constant CSM is not well understood. It may  originate from the interaction of a short lived Wolf-Rayet star wind with the  remnant of a dense red supergiant wind \cite{2014ARA&A..52..487S}, due to  mass loss \cite{2012MNRAS.423L..92Q}, or to the ejection of the H envelope caused by  binary interactions~\citep{Brethauer:2020bmy}. 

\item Model B---An asymmetric CSM model~\citep[][]{2022arXiv220600842B}. The  asymmetry is proposed to be caused by  the H-rich disk in the equatorial plane and the observed X-ray emission from SN 2014C is attributed to this disk like CSM~\citep[][]{2022arXiv220600842B}. Two different density profiles have been proposed for the disk,  one with a density profile falling as $r^{-1.50 \pm 0.01}$ and the other with a steeper profile of the form $r^{-2.42\pm 0.17}$. We take into account both density profiles: Model B1 ($r^{-1.5}$) and Model B2 ($r^{-2.42}$).

To model the asymmetric CSM scenario, we introduce  a geometrical (asymmetry) factor, $f$ ($\leq 1$)\citep[][]{2022arXiv220600842B}.  The case $f=1$ corresponds to  spherical symmetry in the CSM and the most asymmetric (disk-like) CSM is described by $f = 0.1$. The degree of the asymmetry of the CSM of SN 2014C is still uncertain. Therefore, to take into account the possibility of different  asymmetric scenarios, $f$ is varied between $1.0$ and $0.1$~\citep[][]{2022arXiv220600842B}. 
The variation of the CSM density  is proportional to $f$ for Model B1, whereas it scales as $\sqrt{f}$ for Model B2. 
\end{itemize}
 
 We assume that the CSM  ends abruptly at the outer radius ($r_{\rm o}$), for Models A and B. The location of the CSM over density is uncertain \citep[][]{Margutti:2016wyh,2022ApJ...930..150V,2022arXiv220600842B}, hence    we choose to keep the location unchanged in both models. 
Note that these CSM profiles are  different with respect to the conventional wind  one ($\propto r^{-2}$ \citep[][]{Margutti:2016wyh}), not considered in this paper. Here we refer the reader to Refs.~\cite{2022JCAP...08..011S,Murase:2018okz} for dedicated work on the production of neutrinos and gamma-rays for the CSM wind profile.

 \section{Spectral energy distributions of gamma-rays and neutrinos} \label{sec:model}
 High energy neutrinos and gamma-rays can be produced through the interaction of shock accelerated protons with  non-relativistic CSM protons. This proton-proton ($pp$) interaction creates  charged and neutral mesons ($\pi$ and $\eta$), which decay into secondaries, such as neutrinos and gamma-rays \citep[][]{Kelner:2006tc}. 
 
 The spectral energy distribution  of  accelerated protons is assumed to be a power-law distribution, $Q_{\rm p}^{\rm inj} (E_{\rm p},r)\propto E_{\rm p}^{-\alpha} \exp\left[{-E_{\rm p}/E_{\rm p,max}(r)}\right]$, where $\alpha$ is the power law index \citep{1982ApJ...258..790C,1996A&A...305L..53M,Kelner:2006tc, Petropoulou:2016zar,Ofek:2014fua,Petropoulou:2017ymv,Murase:2017pfe}. 
We consider $\alpha \in [2.0, 2.2]$ for our analysis \cite{Murase:2018okz,Blasi:2013rva,2001RPPh...64..429M,1987PhR...154....1B}. The choice of $\alpha$ depends on the details of the shock acceleration mechanism, also responsible for  efficiently accelerating protons up to PeV energies. In particular, magnetic field amplification can be considered to be the primary requirement for efficient acceleration \citep{Cardillo:2015zda,Cristofari:2022kqv}. For example,  plasma instabilities may give rise to  small scale magnetic field~\cite{Murase:2014bfa}. Non-resonant hybrid (NRH)  instability  \citep{10.1111/j.1365-2966.2004.08097.x,10.1111/j.1365-2966.2005.08774.x,10.1046/j.1365-8711.2000.03363.x,10.1093/mnras/stt1371,10.1093/mnras/stt179} in  YSNe is another possibility. Such instability investigated for SN remnants~\cite{10.1111/j.1365-2966.2004.08097.x} shows that cosmic rays (CRs) in the upstream shock can excite turbulence amplifying the initial background magnetic field.  Such amplification can lead to long confinement of CRs allowing for acceleration to very high energies. In the SN remnant environment, the interaction of the strong shock with the upstream CRs is considered to be the requirement for the NRH instability. Similar amplification in YSNe also becomes feasible due to the high shock speed ($\sim 0.1c$) produced by these objects, see e.g.~Ref.~\cite{10.1093/mnras/stt1371} for more details.


The maximum proton energy,  $E_{\rm p,max}(r)$,  governs the shape of the proton spectra at higher energies.  $E_{\rm p,max}(r)$ is determined by balancing the acceleration timescale with the cooling timescales, i.e., $t_{\rm acc}(r)=\min[t_{\rm ad}(r),t_{\rm pp}(r)]$, where $t_{\rm ad}$ and $t_{\rm pp}$ are the cooling timescales  for  adiabatic losses and $pp$ collisions, respectively. 
The acceleration timescale is given by $t_{\rm acc}= 6 E_{\rm p} c/ e B v_{\rm sh}^2$ in the Bohm  limit, where $B$ is the magnetic field strength of the post shock CSM  given by $B=3/2 [4 \pi \epsilon_{\rm B} m_{\rm p} n_{\rm CSM}(r) v_{\rm sh}^2]^{1/2}$~\citep[][]{Petropoulou:2016zar}. The fraction,
 $\epsilon_{\rm B}$, of the post shock thermal energy converted to magnetic energy  \citep[][]{Petropoulou:2016zar} 
 can be estimated from SN radio observations and is typically in the range  $10^{-3}$--$10^{-2}$ \citep[][]{Milisavljevic:2015bli, Margutti:2016wyh,Murase:2018okz,2022JCAP...08..011S,2021ApJ...923L..24S}. The shock velocity, $v_{\rm sh}$, slowly decreases  as a function of the radius, therefore we assume that it is  constant, $\mathcal{O}(10^{4})$ $\rm km\ s^{-1}$~\citep{Ofek:2013afa,Ofek:2014fua,2021ApJ...923L..24S,Margutti:2016wyh,2022ApJ...930..150V,Brethauer:2020bmy,2022arXiv220600842B}.  For a typical LT YSN, with $n_{\rm CSM} \sim 10^6~\rm cm^{-3}$, $v_{\rm sh}\sim 10^4~\rm km~s^{-1}$, $\epsilon_{\rm B} \sim 10^{-2}$,  and $B \sim \mathcal{O}(1)$~G.  This large magnetic field can ensure long confinement of protons in the shocked CSM accelerating them to very high energies; {\color{black} see Appendix~\ref{sec:appendix} and Refs.~\cite{Murase:2010cu,Murase:2014bfa}.} 
 The acceleration timescale for YSNe remains competitive to the different loss timescales.
 In particular, the acceleration of protons may be limited by  cooling process as well as dynamical losses. The cooling processes include inelastic $pp$ interactions and different photo-hadronic interactions such as photopion and photopair production. However, it has been shown for YSNe that photo-hadronic interactions are suppressed due to the low energy of the target photons~\cite{Murase:2010cu,2012IAUS..279..274K,Petropoulou:2016zar,Sarmah:2022vra}. Hence, the only relevant loss timescales are dynamical or adiabatic and the $pp$ collision timescales . The adiabatic timescale is defined as $t_{\rm ad} (r) \sim r/{v_{\rm sh}}$ and  the $pp$ interaction timescale is given by $t_{\rm pp}(r)= [\kappa_{\rm pp} \sigma_{\rm pp} n_{\rm CSM}(r) c]^{-1}$, where $\kappa_{\rm pp}=0.5$ is the proton inelasticity and $\sigma_{\rm pp}$ is the $pp$ interaction cross-section \cite{Kelner:2006tc}. For a typical LT YSN scenario (see Table~\ref{tab:parameters}), these timescales  are $t_{\rm acc} \sim 6 \times 10^{5} (E_{\rm p}/{\rm PeV})$~s, $t_{\rm ad} \sim 6\times 10^7$~s, $t_{\rm pp} \sim 10^{8}$~s. In addition, the diffusion of particles may also affect the acceleration as well as the $pp$ interaction. For a Kolmogorov like diffusion~\cite{Celli:2019qcs}, the diffusion timescale is, $t_{\rm diff} \sim  10^{9}/ \sqrt{E_{\rm p}/{\rm PeV}}~{\rm s} $. This shows that the acceleration timescale for PeV protons is significantly smaller than the relevant loss timescales (see Appendix~\ref{sec:appendix} for details). Thus, the acceleration of protons to PeV energies in a LT YSN environment can be possible due to such short timescale i.e., a few years.  
 
The dependence of the maximum proton energy,  $E_{\rm p,max}(r)$ on the  parameters discussed above can be obtained  from the relation, $t_{\rm acc}(r)=\min[t_{\rm ad}(r),t_{\rm pp}(r)]$. In particular,  $E_{\rm p,max}(r)$ depends on  $\epsilon_{\rm B}$,  $v_{\rm sh}$ and $n_{\rm CSM}$. Larger $v_{\rm sh}$ and $\epsilon_{\rm B}$ are responsible for larger $E_{\rm p,max}(r)$, while a denser CSM slows down the shock, leading to a smaller $E_{\rm p,max}(r)$. 
Other possible losses, such as synchrotron or inverse Compton losses, are negligible, see e.g.~Ref.~\citep{2022JCAP...08..011S}. In addition to these loss timescales, the confinement time of the protons 
needs to be  larger than the acceleration timescale to prevent the particles from escaping the acceleration region. This requires the maximum wavelength of the scattering turbulence ($\lambda_{\rm max}$) to be larger than the gyro-radius ($r_{\rm g}$) of the particles \citep{10.1093/mnras/stw123}. The turbulence could be caused by the interaction of the accelerated protons with the upstream CSM~\citep{10.1111/j.1365-2966.2004.08097.x}. 
However, if $\lambda_{\rm max} \ll r_{\rm g}$, the maximum proton energy, $E_{\rm p,max}(r)$ could be smaller than PeV \citep{Petropoulou:2016zar}.  Hence, the detection of secondary signals (gamma-rays and neutrinos) and their energy will provide crucial information on the acceleration efficiency.

The normalization of the injection proton distribution $Q_{\rm p}^{\rm inj} (E_{\rm p},r)$ depends on  the SN energy budget  going into  protons. The fraction, $\epsilon_{\rm p}$, of the kinetic energy going to the protons is kept as a free parameter and assumed to be in the range of 
$0.01$--$0.1$ \citep[][]{Milisavljevic:2015bli, Margutti:2016wyh,Murase:2018okz,2022JCAP...08..011S}.
The total kinetic energy per unit radius released in the explosion is given by $E_{\rm KE}= (9 \pi/8) m_{\rm p} v_{\rm sh}^2 r^2 n_{\rm CSM}(r)$, where $n_{\rm CSM}(r)$ is the CSM density profile and $m_{p}$ is the proton mass  \citep{Petropoulou:2016zar}.

The steady state proton distribution, $\mathcal{N}_{\rm p} (E_{\rm p},r)$, is obtained from the following equation~\cite{Petropoulou:2016zar}:
\begin{align}
\frac{\partial \mathcal{N}_{\rm p}(E_{\rm p},r)}{\partial r}+\frac{\mathcal{N}_{\rm p}(E_{\rm p},r)}{v_{\rm sh} t_{\rm pp}(r)}-\frac{\partial}{\partial E_{\rm p}}\left[\frac{E_{\rm p} \mathcal{N}_{\rm p}(E_{\rm p},r)}{r}\right] \nonumber \\= \mathcal{Q}_{\rm p}^{\rm inj}(E_{\rm p},r)\ ,
\label{proton_de}
\end{align}
where the second and third terms  take care of the $pp$ interaction and adiabatic losses, respectively. The injection spectra, $Q^{\rm inj}_{\rm i}(E_{\rm i},r) \propto \mathcal{N}_{\rm p} (E_{\rm p},r)$, for the secondary particles are estimated from the steady proton distribution, where  $i= \gamma~{\rm or}~\nu_{\rm f}$ and $f$ is the neutrino flavor, see Ref.~\citep{Kelner:2006tc} for details. Note that  we do not distinguish between neutrinos and antineutrinos. The secondary particles also depend on the escape time  in the CSM environment ($t_{\rm esc} \sim r/ 4 c$), which is  governed by the following equation \citep[][]{Petropoulou:2016zar}:
\begin{equation}
    \frac{\mathrm{d} \phi^{\rm S}_{\rm i}(E_{\rm i},r) }{\mathrm{d}r} + \frac{\phi^{\rm S}_{\rm i}(E_{\rm i},r) }{v_{\rm sh} t_{\rm esc}(r)} = \mathcal{Q}^{\rm inj}_{\rm i}(E_{\rm i},r)\ ,
\end{equation}
where  $\phi^{\rm S}_{\rm i}(E_{\rm i},r) $ is the steady state secondary  spectrum.

The secondary (gamma-rays and neutrinos) flux at Earth from a SN burst at luminosity distance $D_{\rm L}$ is further modified by  loss processes in the source (S) as well as in the intergalactic medium during propagation (P) to Earth. Hence, for a source at redshift $z$, the flux at Earth is:
\begin{align}
    \phi_{\rm i}(E_{\rm i},r)= &\left[ \frac{   e^{-\tau_{loss}^{\rm P}(E_{\rm i}^{\prime})}      }{4 \pi D_{\rm L}^2(1+z)^2 t_{\rm esc}(r)} \right]  \nonumber \\ & \times  \left[ \phi^{\rm S}_{\rm i}(E_{\rm i}^{\prime},r) e^{-\tau_{loss}^{\rm S}(E_{\rm i}^{\prime},r)}\right]\ ,
\label{eq:NuFlux}
\end{align}
where $E_{\rm i}^{\prime}=(1+z)E_{\rm i}$.
Gamma-rays  suffer energy loss from pair production on low-energy thermal photons. The amount of gamma-ray attenuation at the source is determined by the optical depth, $\tau_{loss}^{\rm S}(E_{\rm i}^{\prime},r)$, which depends on the density of thermal photons in the interaction zone (i.e., CSM) and their average energy \citep[][]{2022JCAP...08..011S}. The thermal photons follow a black-body distribution and the density of these thermal photons falls  as $r^{-2}$~\citep{2022JCAP...08..011S}. 
Due to the radial declination of the thermal photon density, $\tau_{loss}^{\rm S}(E_{\rm i}^{\prime},r)$ decreases as a function of the radius.  The attenuation of gamma-rays  during propagation to Earth scales as $e^{-\tau_{loss}^{\rm P}(E_{\rm i}^{\prime})}$ in Eq.~\ref{eq:NuFlux}.
The  photon background includes the Extra-galactic Background Light (EBL) and the Cosmic Microwave Background (CMB).  The amount of energy losses is linked to the EBL and CMB densities as well as  the distance the gamma-rays travel. 

Neutrinos do not suffer losses during propagation, however the neutrino flux  is modified by flavor conversion, hence we consider the flavor ratio $\nu_e:\nu_\mu:\nu_\tau=1:1:1$  at Earth \cite{PhysRevD.90.023010}. Therefore, the neutrino flux for one flavor at Earth is one third of the three-flavor neutrino flux given by Eq.~\ref{eq:NuFlux} (note that we do not distinguish between  neutrinos and antineutrinos). 
 
Interestingly, the secondary particle (gamma-ray and neutrino) production  beyond the maximum radius, $r_{\rm max}$ = $\min[r_{\rm o}, r_{\rm dec}]$, decreases  fast. The deceleration radius $r_{\rm dec}$ corresponds to the radius where the CSM mass ($M_{\rm CSM}$) swept up by the shock equals the ejecta mass ($M_{\rm ej}$) \citep[][]{2022JCAP...08..011S}. For the constant-density shell (Model A), $M_{\rm CSM}=2~{\rm M_{\odot}}$ and $r_{\rm dec}\sim  10^{17}~{\rm cm}$. Since the interaction of the SN 2014C shock with the CSM  is observed up to $2.5 \times 10^{17}~{\rm cm}$ \citep[][]{Brethauer:2020bmy}, we compute the secondary flux up to this radius,  although the flux is expected to decrease  beyond $r_{\rm dec}$ significantly.

 \section{Temporal evolution and energy distribution of  neutrino and gamma-ray signals}
 \label{sec:tem evolution}
 \begin{table*}[]
 \caption{Characteristic model parameters of  SN Ib/c emission, inspired by  observations of SN 2014C. The second column   lists the model parameter typical of the  early phase with a wind density profile ($r^{-2}$) \citep{2022JCAP...08..011S}. The third column represents the parameter values for the LT emission. Uncertainties on the LT parameters are also reported in the fourth column. 
  }
    \centering
    \begin{tabular}{|c|c|c|c|c|}
    \hline
    {\bf Parameters} & {\bf Early phase}  & {\bf Typical value (LT)} & {\bf Uncertainty range (LT)} & {\bf References}\\
    \hline
    \hline
    $v_{\rm sh}~(\rm km\ s^{-1})$ & $2 \times 10^4$ & $10^4$ & $(4$--$45)\times 10^{4}$ & \cite{Margutti:2016wyh,Brethauer:2020bmy,2022JCAP...08..011S}\\  
    \hline 
    $r_{\rm i}~(\rm cm)$ & $3 \times 10^{11}$ & $6\times 10^{16}$ & $(5.5$--$6) \times 10^{16}$ & \cite{Margutti:2016wyh,Brethauer:2020bmy,2022JCAP...08..011S}\\
    \hline 
    $r_{\rm o}~(\rm cm)$ & $6 \times 10^{16}$ & $2.5 \times 10^{17}$ & $(1$--$2.5) \times 10^{17}$ & \cite{Margutti:2016wyh,Brethauer:2020bmy,2022JCAP...08..011S}\\
    \hline
    $n_{\rm CSM}~(\rm cm^{-3})$ & $2\times 10^{12}$ & $2\times 10^6$ & --- &\cite{Margutti:2016wyh,Brethauer:2020bmy,2022JCAP...08..011S}\\ 
    \hline
    $\epsilon_{\rm p}$ & $10^{-1}$ & $5 \times 10^{-2}$ & $10^{-2}$--$10^{-1}$ & \cite{Milisavljevic:2015bli, Margutti:2016wyh,Murase:2018okz,2022JCAP...08..011S}\\
    \hline
    $\epsilon_{\rm B}$ & $10^{-2}$ & $1.5 \times 10^{-2}$ & $10^{-3}$--$10^{-2}$ & \citep[][]{Milisavljevic:2015bli, Margutti:2016wyh,Murase:2018okz,2022JCAP...08..011S,2021ApJ...923L..24S}\\
    \hline
    $D_{\rm L}$ (Mpc) & 14.7 & 14.7 & $14.1-15.3$ & \cite{2022arXiv220600842B}\\
    \hline
    Onset time  & $180$~s & $250$~days & $(100$--$400)$~days & \cite{Milisavljevic:2015bli,Margutti:2016wyh,Brethauer:2020bmy,2022arXiv220600842B, Tinyanont:2019qol,2022JCAP...08..011S}\\
    \hline
    Declination & \multicolumn{1}{c}{\ \ \ \ \ \ \ \ \ \ \ \ \ \ \ \ \ \ \ ~~~~~~~{$34^{o}$}} & & ---&  \cite{CDS} \\
    \hline
    \end{tabular}
    \label{tab:parameters}
\end{table*}
 Because of the LT CSM interaction, we expect copious production of gamma-rays and neutrinos  about a year after the SN explosion, and   
 the  time evolution of the neutrino and gamma-ray signals should carry  crucial information about the CSM properties.
 For the calculation of the  gamma-ray and neutrino fluxes,  
 our  choice of the benchmark SN model parameters  is motivated by the observations of SN 2014C \cite{Margutti:2016wyh,Brethauer:2020bmy,Tinyanont:2019qol} and summarized in Table~\ref{tab:parameters}. 
 Note that the parameters in Table~\ref{tab:parameters} are the ones common to  all CSM models introduced in Sec.~\ref{sec:CSM}; the  differences among the  models are due to the radial evolution of the CSM density profile between $r_{\rm i}$ and $r_{\rm o}$ and  the asymmetry parameter $f$.

\begin{figure*}
    \centering
    \includegraphics[width=0.7\textwidth]{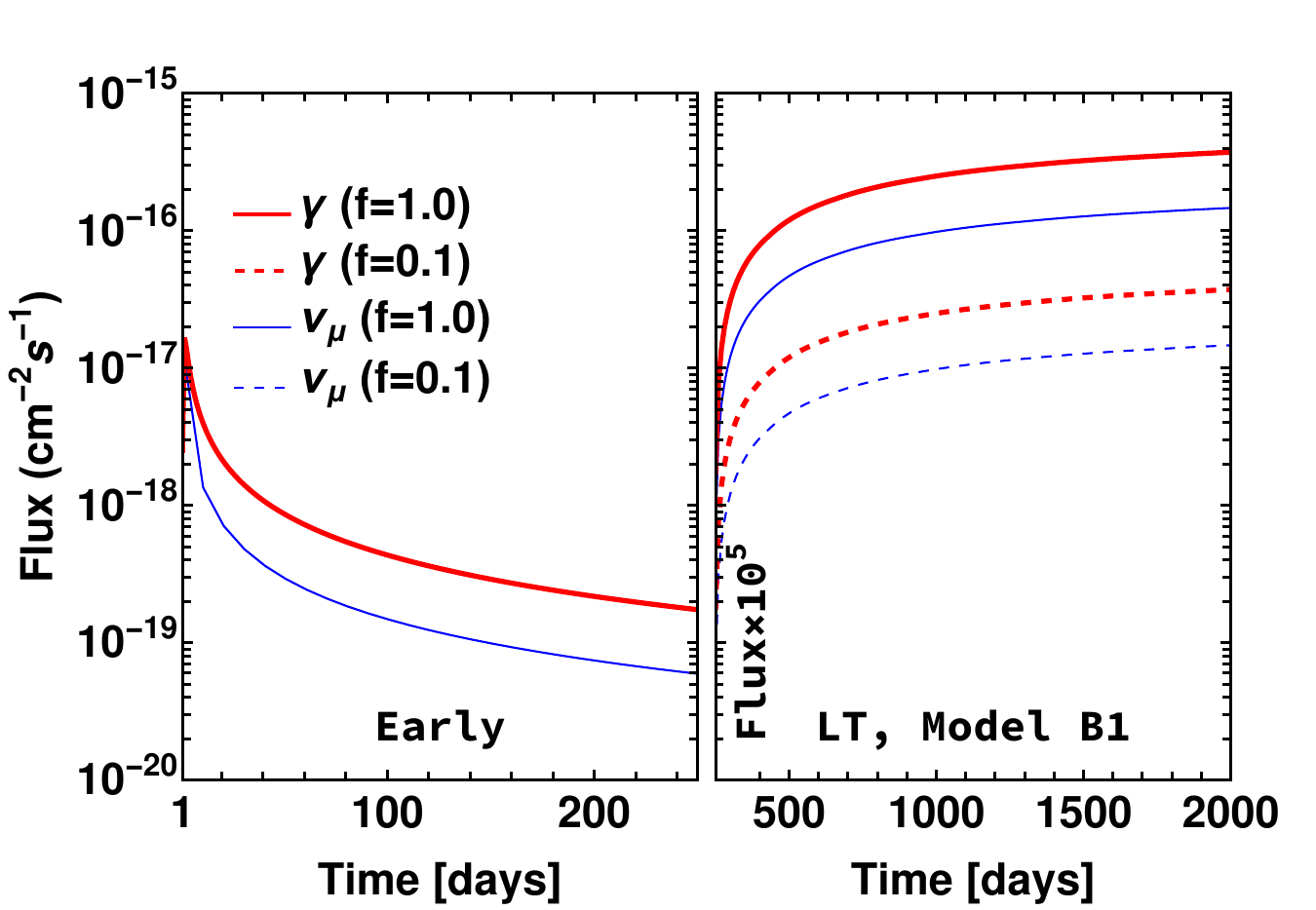}
 \caption{Temporal evolution of the flux of muon neutrinos (thin blue) and gamma-rays (thick red) at Earth from  SN Model B1 (see also Table \ref{tab:parameters}).  The left plot extends up to $250$~days and it shows the early emission from  SN Ib/c; the right panel shows the LT emission, i.e.~beyond  $250$ days. The y-axis of the right panel is rescaled by a factor $10^5$ at $250$ days. The continuous and  dashed line styles represent the minimum ($f=1$) and maximum ($f=0.1$) CSM asymmetry, respectively.   Note that the gamma-ray distributions do not take into account absorption.  One can see that the early time emission from  SNe Ib/c is  significantly smaller than the corresponding LT emission. The other SN models (A and  B2)  show a similar temporal evolution and are therefore not shown.
 } 
    \label{fig:density}
\end{figure*}

Figure~\ref{fig:density} shows   the temporal evolution of the flux of gamma-rays (thick red) and muon neutrinos (thin blue) at Earth  for SN Model B1 ($r^{-1.5}$ profile). 
The continuous and dashed lines show the cases of minimum ($f=1$) and maximum ($f=0.1$) asymmetry of the CSM, respectively \citep[][]{2022arXiv220600842B}. 
The initial emission (up to $250$~days, left panel) is  small as the CSM for  SNe Ib/c is  thin \citep[][]{2022JCAP...08..011S,Murase:2017pfe}. The sharp rise (note the re-scaling of the y-axis at $250$~days, right panel) in the gamma-ray and neutrino spectra  is due to the dense H-rich CSM.  These fluxes are computed  up to $2000$~days that correspond to the outer radius $r_{\rm o}$ of the CSM.

 \begin{figure*}[]
    \centering
    \includegraphics[width=0.47\textwidth]{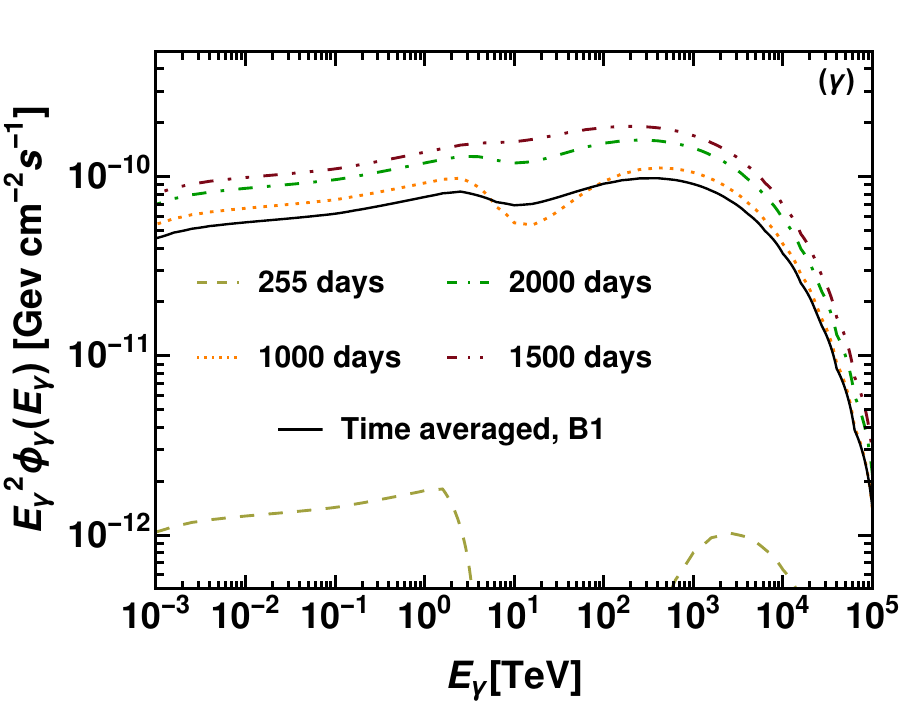}~
     \includegraphics[width=0.47\textwidth]{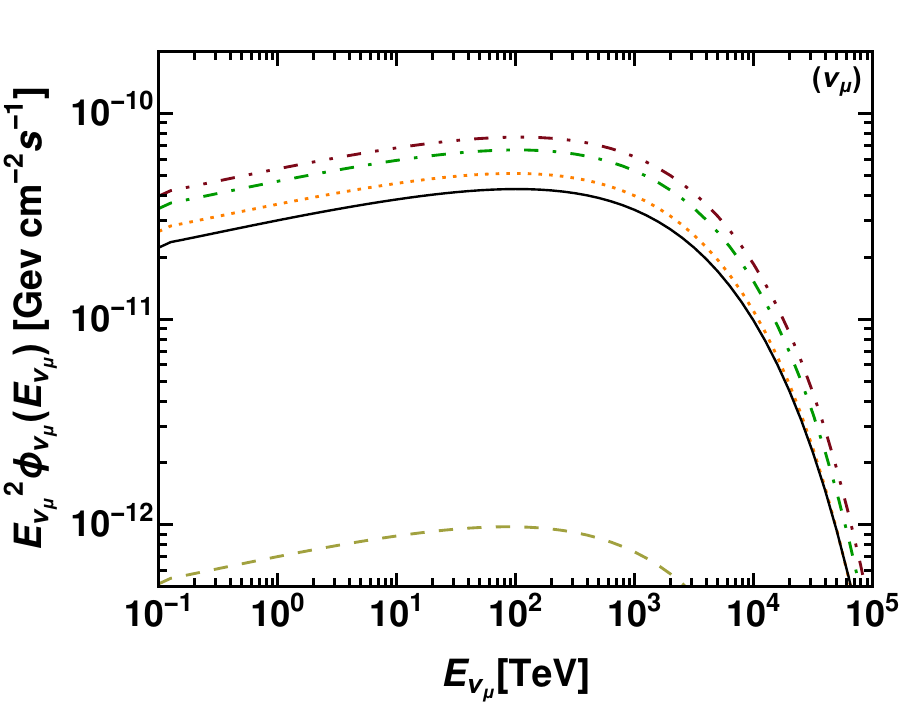} ~
      \includegraphics[width=0.47\textwidth]{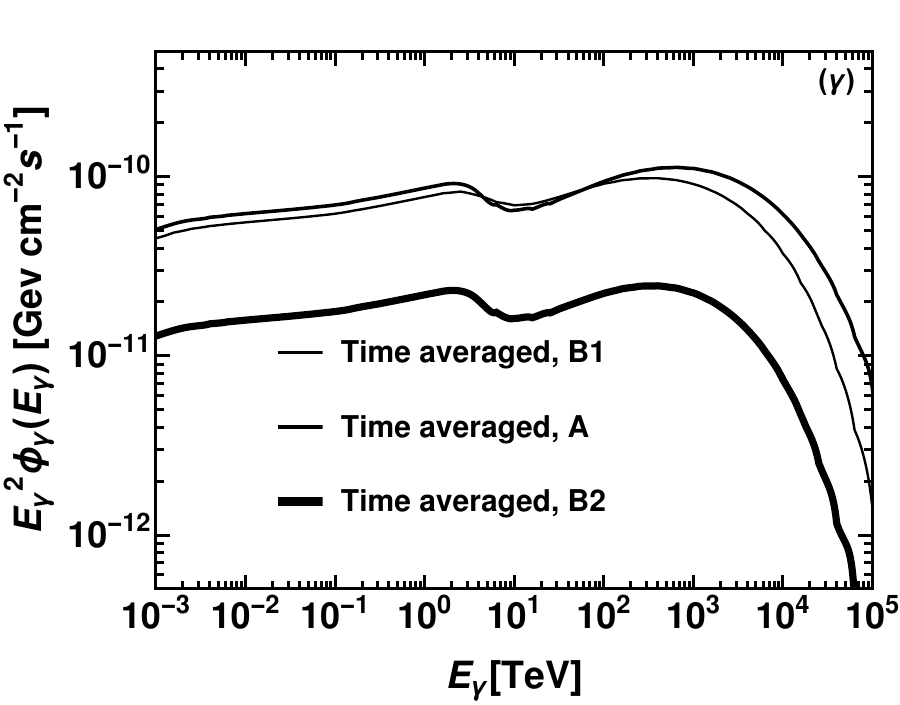} ~
       \includegraphics[width=0.47\textwidth]{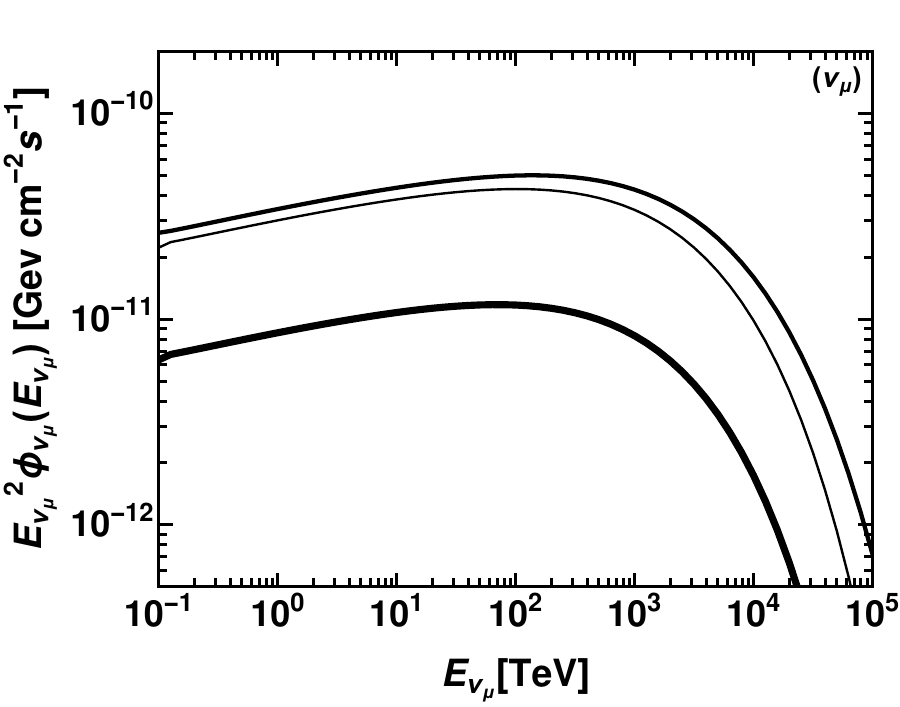} ~
    \caption{ {\it Top:} Time snapshots of   gamma-ray (on the left) and muon neutrino (on the right) fluxes as  functions of the  energy  for SN Model B1 (see  Table \ref{tab:parameters} for more details on the model parameters). The different curves indicate the flux at $255$, $1000$, $1500$, and $2000$ days. We consider the asymmetry factor to be $f=0.5$ to describe the typical spectral variation over the late emission phase. 
    After the onset  of the LT interaction, the fluxes   increase with time. The fluxes averaged over $2000$ days has also been shown by the black  curves. The gamma-ray fluxes between $1$--$10^3$~TeV are attenuated by pair production on the SN thermal photons. The amount of absorption is initially large, but it becomes smaller with time as the thermal photon density falls rapidly with the radius.  {\it Bottom:}  The fluxes of gamma-rays (left) and neutrinos (right) averaged over $2000$ days for Model A (medium thick), B1 (thinnest), and B2 (thickest) are  shown for guidance. 
    } 
    \label{fig:fluxes}
\end{figure*}
 The  energy fluxes for SN Model B1, plotted in the top panel of Fig.~\ref{fig:fluxes} for gamma-rays (on the left) and muon neutrinos (on the right),  reveal the dependence of the production mechanism on the SN model parameters. The  curves in different colors and line styles  represent fluxes at different time snapshots ($255$, $1000$, $1500$, and $2000$ days), highlighting the flux variation over the LT phase. This panel also shows the flux averaged over $2000$ days (black  curve).  
Contrasting the flux at $255$~days (corresponding to the onset of the shock-CSM interaction) with  the one above $1000$~days, one can see that the flux tends to increase with time. 
  The maximum proton energy, $E_{\rm p,max}(r)$, fixes the spectral shape at higher energies as it acts as an exponential cut-off (see Sec.~\ref{sec:model}). Hence, the fluxes of both gamma-rays and neutrinos  fall rapidly above $10^3$ TeV. 

  The gamma-ray fluxes  in  the top left panel of Fig.~\ref{fig:fluxes} 
 include  absorption effects. In order to estimate the amount of absorption,  the average energy and luminosity of thermal photons are assumed to be $0.05$ eV and $5\times 10^{40}~\rm erg/s$ \citep[][]{Brethauer:2020bmy,Margutti:2016wyh}.   
  The gamma-ray fluxes  show dips of different sizes due to  pair production losses on the ambient thermal photons. The dips have different sizes as the optical depth $\tau_{loss}^{\rm S}(E_{\gamma}^{\prime},r)$ falls with the radius \citep{2022JCAP...08..011S};  this implies that the gamma-rays produced at larger radii  have smaller attenuation. However, the  attenuation  during propagation due to the EBL is not significant since SN 2014C is at $14.7$~Mpc \citep[][]{2022JCAP...08..011S}; therefore  we neglect this  effect.     
  
The bottom panel of Figure~\ref{fig:fluxes}  shows the fluxes averaged over $2000$ days for Model A, B1, and B2 (medium thick, thinnest and thickest, respectively). It is important to note that the fluxes for  Model B depend on the  CSM asymmetry factor $f$. The spectral shape  remains the same for different $f$, but the  normalization changes. For example, if we increase $f$ to 1, the time-averaged flux of Model B1 would be larger than the ones of the  other CSM models. Note that the time-averaged fluxes of gamma-rays and neutrinos (black curves) are smaller than the maximum fluxes (at $2000$ days) by a few factors. Hence,  in the following,  we consider the time-averaged fluxes to be conservative estimates of the detection prospects of SN 2014C-like events.

 \section{Detection prospects  for  SN 2014C-like bursts}\label{subsec:results}
 \label{sec:detection prospects}
 In this section, we explore the detection prospects of gamma-rays and  neutrinos from  SNe Ib/c LT with current and upcoming gamma-ray (Fermi-LAT and CTA) and high energy neutrino (IceCube and IceCube-Gen2)  detectors \citep[][]{IceCube:2016tpw,IceCube:2018ndw,IceCube-Gen2:2020qha,KM3NeT:2018wnd,Ambrogi:2018skq}. For comparison, we  consider Model B1 and Model B2 with  CSM asymmetry $f=0.5$.

 \subsection{Current and future detection prospects}

\begin{figure*}
    \centering
      \includegraphics[width=0.47\textwidth]{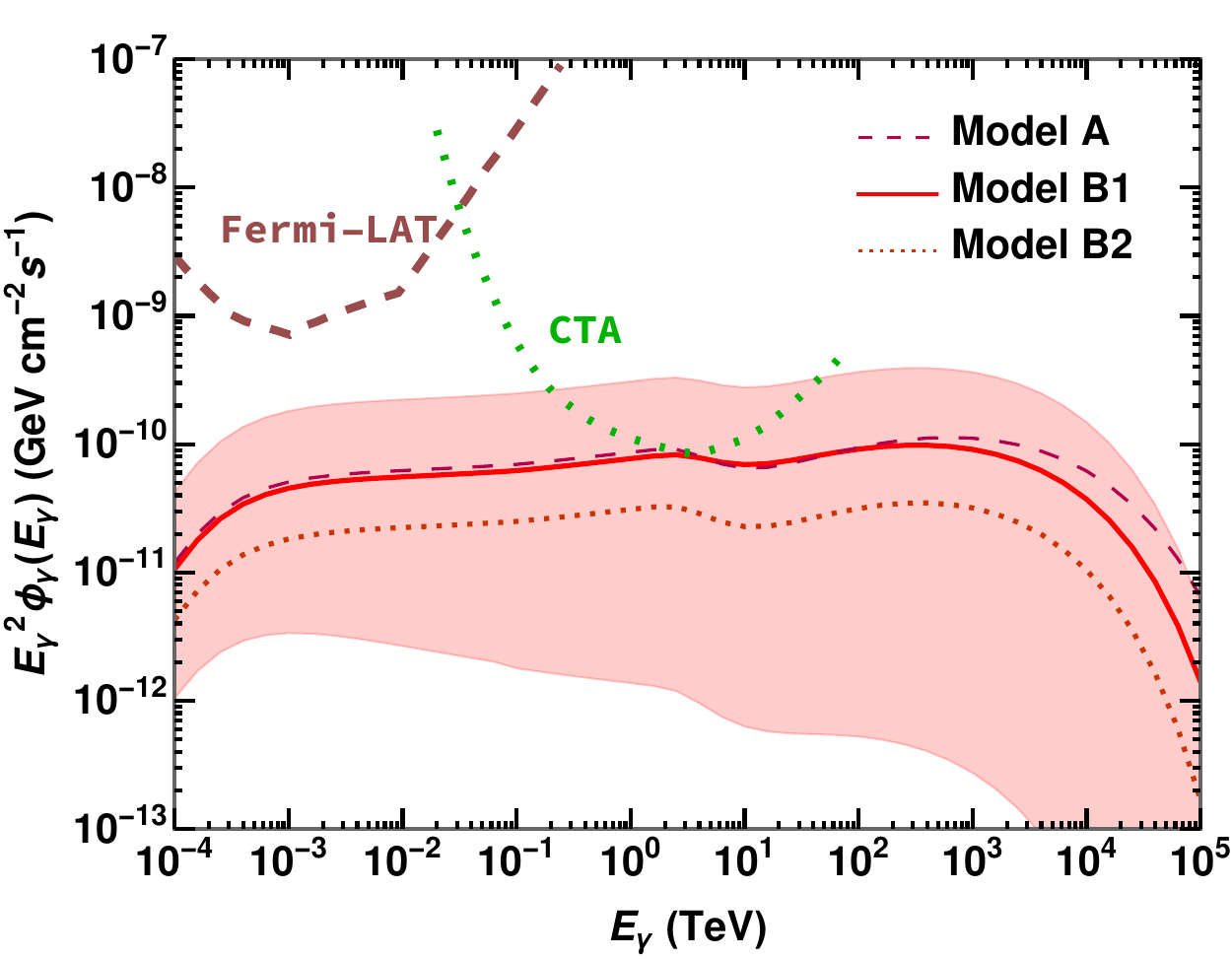}~
      \includegraphics[width=0.48\textwidth]{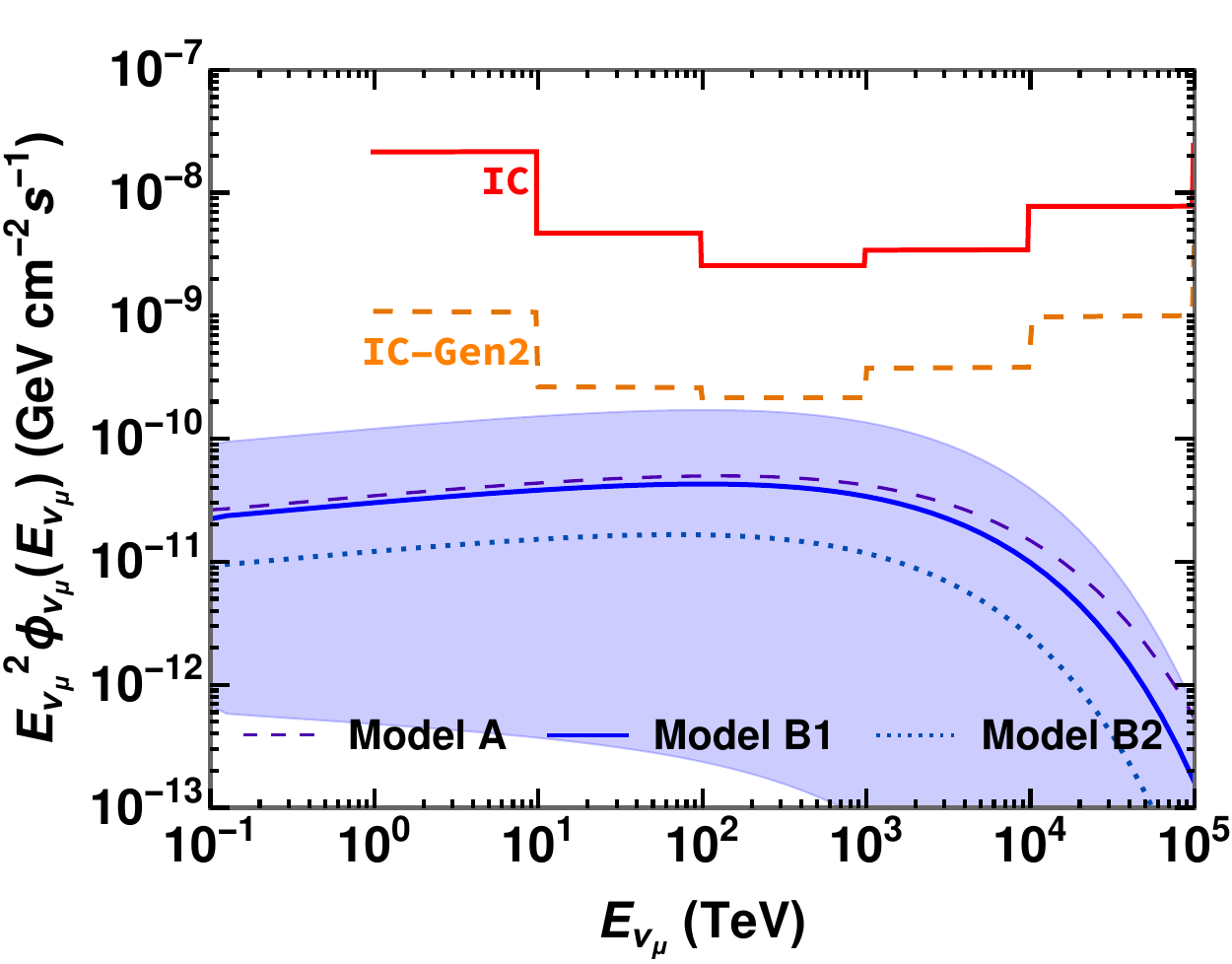}
    \caption{{\it Left:} Time averaged  gamma-ray flux for different CSM models  (see  Table \ref{tab:parameters}). 
    The red band represents the time-averaged flux of  SN Ib/c LT obtained  considering the uncertainties in the model parameters ($\epsilon_{P}\sim 10^{-2}$--$10^{-1}$, $\epsilon_{B} \sim 10^{-3}$--$10^{-2}$ and $f \sim 0.1$--$1$), while the other model parameters are kept fixed as detailed in Table~\ref{tab:parameters}. The uncertainty band as been obtained as follows: for the upper limit, we take $\alpha=2.0$, $\epsilon_{\rm p} =10^{-1}$, $\epsilon_{\rm B} =10^{-2}$ and $f=1$ for Model B1;  for the lower limit, we choose $\alpha=2.0$, $\epsilon_{\rm p} =10^{-2}$, $\epsilon_{\rm B} =10^{-3}$, and $f=0.1$ for Model B2.  The red continuous and  dotted curves show the fluxes for Model B1 and Model B2, respectively, for $f=0.5$. The pink dashed curve represents the flux for Model A. The sensitivities of Fermi-LAT (4 years) and CTA (50 hours) are plotted in brown and green, respectively.  CTA may be able to detect gamma-rays from SNe closer than $\simeq 15$~Mpc. {\it Right:} Corresponding neutrino flux and  sensitivities of  IceCube (6 years with $90\%$ CL) \citep{IceCube:2018ndw}, IceCube-Gen2 (6 years with $90\%$ CL) \citep{IceCube-Gen2:2020qha}.
    The blue band represents the uncertainty in the model parameters. The sensitivity chosen for IceCube corresponds to the one at the declination of  $0^\circ$  where IceCube is  most sensitive.   Neutrinos from  SNe Ib/c LT may be detectable for bursts occurring closer than $15$~Mpc. } 
    \label{fig:Events}
\end{figure*}

The  left (right) panel of  Fig.~\ref{fig:Events} shows the gamma-ray (neutrino) flux for different models of the CSM, as well as  the detection sensitivity of  gamma-ray (neutrino) telescopes. The sensitivity curves of Fermi-LAT and CTA shown in the left panel correspond to  $4$ years and $50$ hours of observation time, respectively. Whereas we show the  $6$ year sensitivities of the neutrino detectors in the right panel. 
The SN model parameters are plagued by various uncertainties. In order to take this into account, we consider a range of variability for the microphysical parameters that contribute to the largest uncertainty in the expected fluxes;  we take $\alpha$,  $\epsilon_{\rm p}$ and $\epsilon_{\rm B}$ to vary in the range $2.0$--$2.2$, $10^{-2}$--$10^{-1}$ and $10^{-3}$--$10^{-2}$, respectively \citep[][]{Milisavljevic:2015bli, Margutti:2016wyh,Murase:2018okz,2022JCAP...08..011S}. 
The shaded bands in both panels of  Fig.~\ref{fig:Events} also take into account  the uncertainties on the CSM profile and the asymmetry factor $f$. Note that  the conventional wind profile ($r^{-2}$) with our benchmark parameters (Table \ref{tab:parameters}) leads to fluxes similar to the ones of Model B1 with $f=0.5$ \cite{2022JCAP...08..011S}.

While the detection prospects are less optimistic for Fermi-LAT, CTA may detect  gamma-rays, if the CSM has a smaller asymmetry (i.e., $f \geq 0.5$) compared to the asymmetry for $f = 0.5$. On the other hand, the non-detection of  gamma-rays with CTA  may contribute to constrain the CSM asymmetry factor $f$. 
The forecasted neutrino flux  is beyond  reach for IceCube (6 years with 90\% confidence level (CL), \citep{IceCube:2018ndw}), in agreement with the fact that SN 2014C was not detected in neutrinos---see e.g.~Ref.~\cite{Abbasi:2023enx}.  
However, IceCube-Gen2  (6 years with 90\% CL~\citep{IceCube-Gen2:2020qha}) will have a better sensitivity and be closest to the predicted flux. For bursts occurring at closer distances than 15 Mpc, IceCube-Gen2 will have a reasonable prospect of detection. 
Interestingly, KM3NeT/ARCA (10 years with 90\% CL, \citep{Ambrogi:2018skq}) is expected to hold similar sensitivity as IceCube-Gen2  (6 years with 90\% CL, \citep{IceCube-Gen2:2020qha}) and may probe similar  SNe Ib/c LT objects.
However, 
due to the limited availability  of the  KM3NeT/ARCA sensitivity for the energy bins over the point source observation time, it is difficult to make a precise estimate of the detector response and 
therefore we choose to do not explicitly show the detection prospects of KM3NeT/ARCA in  Fig.~\ref{fig:Events}.  
Also, any comparison of the detector sensitivities with the predicted neutrino fluxes only provides  a broad idea about the detection prospects. This is because of the different sensitivities at different energy bins~\citep{2011ApJ...732...18A,IceCube:2018ndw}. For a robust forecast, one should  compute the number of events considering the impact of the backgrounds. Since our neutrino flux prediction has large astrophysical uncertainties, we only investigate the differential flux  sensitivities of the detector to give an idea of the detection prospects. 

The  secondary fluxes for the asymmetric CSM models
 are computed by assuming that  the disk (which gives rise to the asymmetry in the CSM) is aligned with the observer's line of sight. 
 When the disk is not along the  line of sight, the fluxes might be smaller than the ones shown in Fig.~\ref{fig:Events}. However,  this uncertainty  lies within the uncertainty bands shown in Fig.~\ref{fig:Events}.

\subsection{Detection horizon}

The detection prospects of SNe with LT emission  depend on the rate of such events. In the local universe, we expect about $26\%$  SNe Ib/c~\cite{2014ARA&A..52..487S}, of these about $10\%$ should be Ib/c LT \cite{Margutti:2016wyh}. Thus, the local rate of SNe Ib/c LT is about $2.6\%$ of the  local core collapse SN rate ($1.25 \pm 0.5 \times 10^{-4} ~\rm Mpc^{-3}yr^{-1}$~\citep[][]{Lien:2010yb}).

To  investigate upcoming  detection prospects, we consider  SN 2014C as the benchmark SN Ib/c LT  (Table \ref{tab:parameters}, third column)  and   calculate  the SN detection horizon defined  as the distance at which the source should be located,   for which the energy integrated  flux (averaged  over  $2000$~days) falls below the telescope sensitivity. The energy range for these integrated fluxes is optimized according to the telescope sensitivity. 
For Fermi-LAT and CTA, we consider $10^{-4}$--$10^{-2}$~TeV and  $5 \times 10^{-2}$--$5 \times 10^{1}$~TeV respectively, whereas  we focus on   $10^2$--$10^4$~TeV   for  the neutrino telescopes.    
Note that the sensitivity for the neutrino telescopes depends on the  declination \cite{IceCube:2016tpw,IceCube:2018ndw,IceCube-Gen2:2020qha}. Hence, unlike the gamma-ray telescopes, we consider a maximum and minimum sensitivity resulting in a band. 

\begin{figure*}
    \centering
    \includegraphics[width=0.47\textwidth]{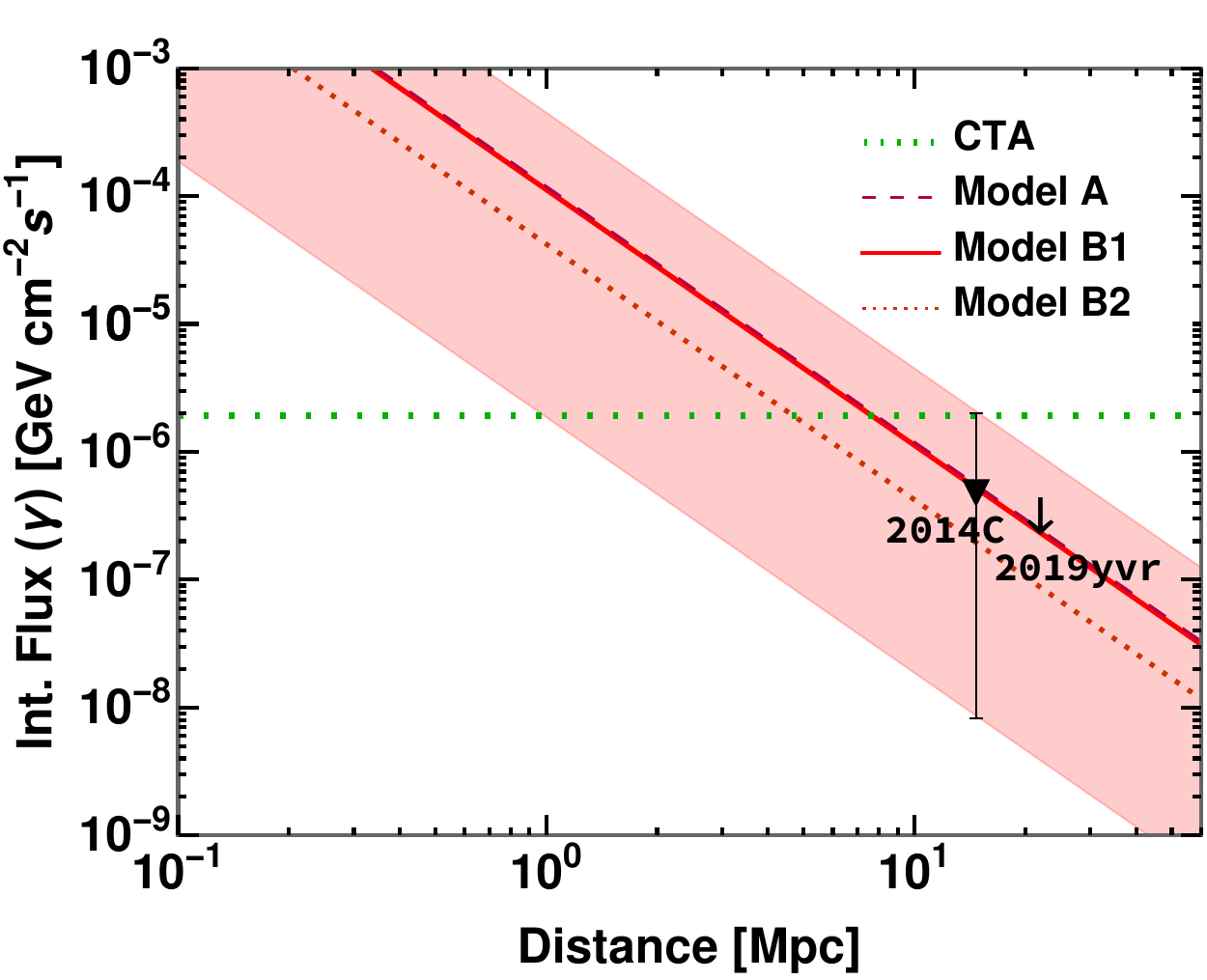}~
    \includegraphics[width=0.48\textwidth]{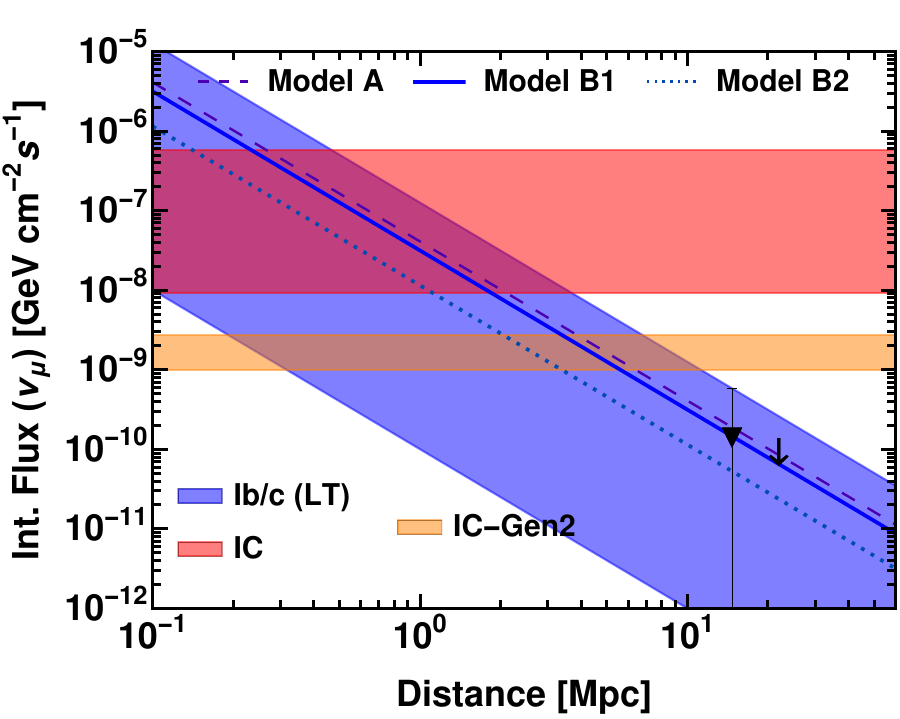}
    \caption{ {\it Left:} Detection horizon (see main text for details) in gamma-rays  for  SNe Ib/c LT.  The time-averaged  gamma-ray flux has been integrated over energy in the range $5 \times 10^{-2}$--$5 \times 10^{1}$~TeV for  SN Model B1 (red) and {Modlel B2} (red dotted) for an asymmetry factor $f=0.5$. The pink dashed line represents the flux for SN Model A,  slightly overlapping with the SN Model B1 flux. The red band corresponds to the uncertainties in the parameters  $\alpha\in [2.0-2.2]$, and $\epsilon_{P}\in [10^{-2}, 10^{-1}]$ and $\epsilon_{B} \in [10^{-3}, 10^{-2}]$, and $f \in [0.1, 1]$. The horizontal dotted line represents the CTA sensitivity. The detection horizon is about $10$~Mpc.  {\it Right:} Corresponding detection horizon for neutrinos.  The blue band represents the time-averaged  flux integrated in the range $10^2$--$10^4$~TeV. The purple dashed, blue and light blue dotted lines show the fluxes for the SN Model A,  Model B1 and Model B2 respectively. The $90\%$ CL sensitivity bands of different detectors are obtained by considering the variation of the declination angle ($\delta$) of the source. The band for IceCube corresponds to the minimum detector  sensitivity for $\delta=-60^\circ$ \citep{IceCube:2016tpw} and maximum detector sensitivity, $\delta=0^\circ$  \citep[][]{IceCube:2018ndw}.  Similarly, we consider $\delta=30^\circ$ (minimum) and $\delta=0^\circ$ (maximum) for IceCube-Gen2~\citep{IceCube-Gen2:2020qha};  
    IceCube-Gen2  has the potential to detect  SNe Ib/c LT up to $10$~Mpc, while IceCube can only probe SNe up to  $3$~Mpc. For guidance, the black data points in both panels show the fluxes of recent nearby SN Ib/c LTe: SN 2014C and SN 2019yvr.}
    \label{fig:horizon_gamma}
\end{figure*}

Figure~\ref{fig:horizon_gamma} shows the detection horizon  for gamma-rays (left) and neutrinos (right). The left panel only shows the detection horizon of CTA;  Fermi-LAT is not shown  because of its weak sensitivity  and different energy range compared to CTA (see Fig.~\ref{fig:Events}). The gamma-ray fluxes for Model B1 and {Model B2} are represented by red  continuous and dotted curves respectively, while the pink dashed line shows the flux for Model A. The shaded bands in both panels correspond to the uncertainty in the parameters ($\alpha \in [2.0-2.2]$, $\epsilon_{P} \in [10^{-2}, 10^{-1}]$ and $\epsilon_{B} \in [10^{-3}, 10^{-2}]$) and CSM asymmetry factor ($f \in [0.1, 1]$) as in Fig.~\ref{fig:Events}.  The horizontal dotted line represents the sensitivity of CTA.  The detection horizon for CTA extends up to $10$~Mpc, while the detection horizon of Fermi-LAT is limited to  $4$~Mpc (results not shown here). Interestingly, the density profile of the CSM for SN (Model B1 and Model B2) plays an important role in the detectability of such SNe. For example, the detection horizon of CTA is about $10$~Mpc for Model B1 and about $6$~Mpc for Model B2. 

The detection horizons of current and upcoming neutrino telescopes are shown in the right panel of Fig.~\ref{fig:horizon_gamma}. The blue continuous and light-blue dotted lines show the integrated $\nu_{\mu}$ flux as a function of the SN distance (Mpc)  for SN Model B1 and Model B2, respectively, and the purple dashed line corresponds to Model A. 
The blue band takes into account the model uncertainties (\PS{$\alpha$}, $\epsilon_{\rm p}, ~\epsilon_{\rm B}, ~f $). 
The upper and lower limits of the  sensitivity of neutrino telescopes depend on the SN declination angle and are shown as  bands in Fig.~\ref{fig:horizon_gamma}. 
The most optimistic model prediction (upper limit of blue band) and most sensitive future telescopes (lower limit of orange and green bands) combination imply that   SNe Ib/c LT may be detected up to  $10$~Mpc with IceCube-Gen2. 
On the other hand, the detection horizon of IceCube (red band) is limited to about $4$~Mpc. 

For guidance, we also show in Fig.~\ref{fig:horizon_gamma} the gamma-ray and muon neutrino fluxes of  SN 2014C that occurred at $14.7$~Mpc as well as the ones of SN 2019yvr observed at $22$~Mpc \cite{Kilpatrick:2021hup}.
The flux of SN 2014C (black inverted triangle) is obtained  by relying on the same parameters as the ones of  the blue line. As for SN 2019yvr, we have chosen  $r_{\rm i} = 3\times 10^{16}~\rm cm$ \citep{sun2022environmental,Kilpatrick:2021hup}, $\epsilon_{\rm p} =0.1$, and $\epsilon_{\rm B}=0.01$ to compute the flux upper limits.  The other parameters for SN 2019yvr are the same as  in Table \ref{tab:parameters}, due to limited information otherwise available for them.  Both these events lie beyond the detection horizon of gamma-ray and neutrino telescopes.  These findings are in agreement with the  non-observation of neutrinos with IceCube and gamma-rays with Fermi-LAT from SN 2014C and SN 2019yvr. 
Similar LT shock-CSM interaction has been observed in  other SNe Ib/c, such as SN 2003gk (estimated distance: $45$~Mpc \cite{2014MNRAS.440..821B}),  SN 2004dk (estimated distance: $21.05$~Mpc \cite{Mauerhan:2018wes}), SN 2004cc (estimated distance: $18$~Mpc \cite{2012ApJ...752...17W}), and SN 2004gq (estimated distance: $26$~Mpc  \cite{2012ApJ...752...17W}). 
Due to their larger or comparable distances, we do not include them  in Fig.~\ref{fig:horizon_gamma}, since we expect comparable or worse detection prospects. 

In Ref.~\citep[][]{2022JCAP...08..011S}, the discovery horizon of IceCube-Gen2 for SN 2014C-like events was found to be about $6$~Mpc. However, here we report  a detection  horizon of  $10$~Mpc. This is because the results in Ref.~\citep[][]{2022JCAP...08..011S} were based on a wind-like CSM profile ($r^{-2}$) whereas  we consider  different CSM profiles in this paper (see Sec.~\ref{sec:CSM}). In addition, the sensitivity of IceCube-Gen2 corresponding to $5\sigma$ CL considered in Ref.~\citep[][]{2022JCAP...08..011S}  is smaller than the $90\%$ CL sensitivity considered in this work.
This also holds for the detection horizons of IceCube. 
The discovery horizon of CTA for SN 2014C-like events was found  to be about $2$--$6$~Mpc in Ref.~\cite{Murase:2018okz}  depending on the energy of gamma-rays and considering the SN emission up to $396$~days, for a dense wind-like CSM~\cite{Margutti:2016wyh}. These conclusions are in agreement with our findings.

\section{Conclusions}
\label{sec:conclusion}

     Late X-ray data of some SNe of Type Ib/c, such as SN 2014C,  have revealed the presence of a dense hydrogen rich CSM far away from the stellar core, whose origin is not yet well understood and still subject of investigation. High energy protons accelerated in the SN shock and interacting with the CSM can lead to the production of secondary particles, such as  gamma-rays and  high energy neutrinos. Yet, the emission of these high energy particles strongly depends on the efficiency of the acceleration mechanism. The acceleration efficiency  would be suppressed  if the shock-CSM interaction fails to produce sufficient turbulence and magnetic field amplification. Hence, it is crucial to look for neutrino and gamma-ray signals to assess the acceleration efficiency.

     In this paper, we have computed the fluxes of  gamma-rays and  high energy neutrinos from  SNe Ib/c LT, considering  SN 2014C as the prototype  SNe Ib/c with LT emission. Because of the uncertainties related to the properties of the CSM,  we have considered three different CSM models: Model A (symmetric, $r^{-2.5}$), Model B1 (asymmetric, $r^{-1.5}$) and Model B2 (asymmetric, $r^{-2.4}$). According to the CSM profile, we predict a range of variability for the expected  fluxes of neutrinos and gamma-rays.

     Based on the observation of SN 2014C and the uncertainties in the model parameters, 
     we have  investigated present and  future detection prospects of SNe Ib/c LT in neutrinos and gamma-rays. 
     We find that the detection horizon for Fermi-LAT and CTA is $4$~Mpc and $10$~Mpc, respectively. Similarly, for neutrinos, the detection horizon of IceCube is about $4$~Mpc, while IceCube-Gen2 can potentially detect SNe Ib/c LT up to  $10$~Mpc. However, the detection horizon in neutrinos can vary up to  a few Mpc depending on the source declination, because of the related neutrino telescope sensitivity. 
     The highly symmetric CSM models are found to  have the best detection prospects while increasing the asymmetry in the CSM  worsens the detection prospects. 
     Our findings are in agreement with the non-detection of gamma-rays and neutrinos from SN 2014C and SN 2019yvr and other SN bursts with LT emission occurring at larger distances. Yet,  upcoming detection of neutrinos and gamma-rays from local SNe Ib/c LT will be crucial  to probe  the CSM properties and the nature of such transients.

    The modeling of the gamma-ray and neutrino emission from  SNe Ib/c LT presented in this work is based on  the observations of  SN 2014C. 
    Considering the  frequency of SN 2014C-like events in the recent past, one might expect to shed light on the properties of the CSM of  SNe Ib/c exhibiting LT emission with upcoming radio and X-ray observations.   In addition,  future gamma-ray and neutrino telescopes will provide complementary information, if SN bursts with LT emission should  occur within $15$~Mpc.

\acknowledgements
 S.C. acknowledges the support of the Max Planck India Mobility Grant from the Max Planck Society, supporting the visit and stay at MPP during the project. S.C has also received funding from DST/SERB projects CRG/2021/002961 and MTR/2021/000540. I.T.~thanks the Villum Foundation (Project No.~37358), the Carlsberg Foundation (CF18-0183) for support, as well as the Deutsche Forschungsgemeinschaft through Sonderforschungbereich SFB 1258 ``Neutrinos and Dark Matter in Astro- and Particle Physics'' (NDM). K.A.~was supported by the Australian Research Council Centre of Excellence for All Sky Astrophysics in 3 Dimensions (ASTRO 3D), through project number CE170100013.

\appendix
\label{sec:appendix}
\section{Timescales for proton acceleration and different cooling processes}

\begin{figure}
    \centering
    \includegraphics[width=0.49\textwidth]{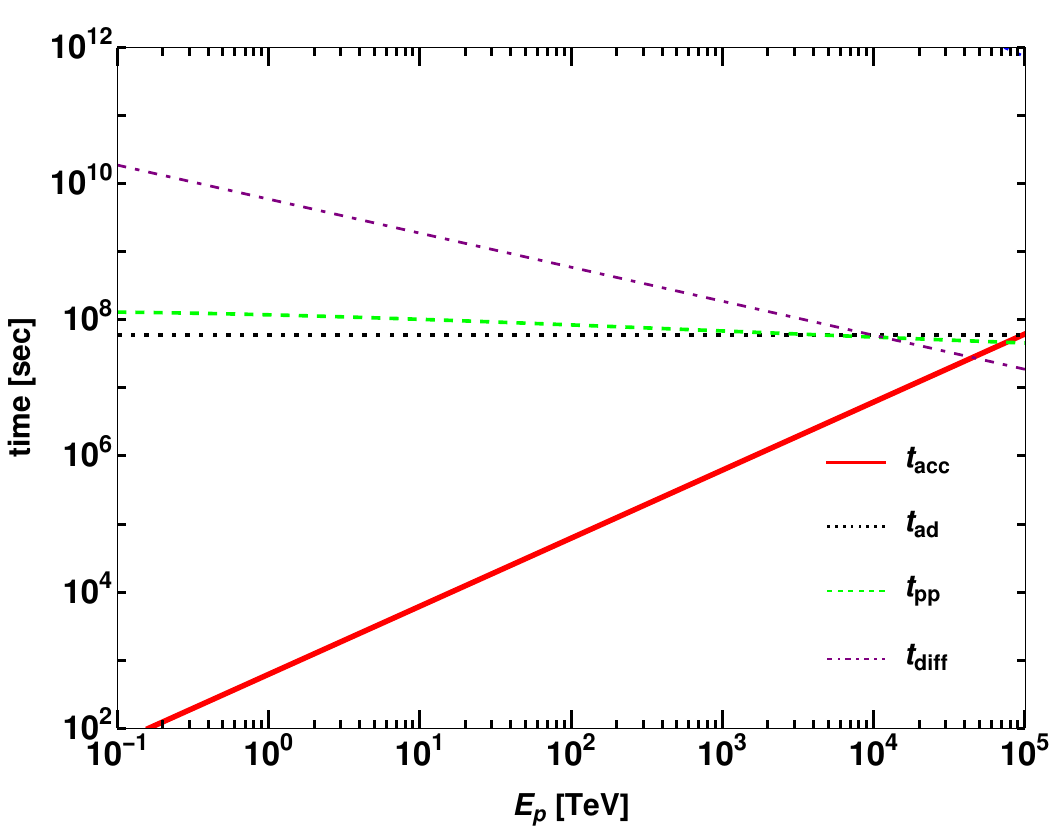}
    \caption{Acceleration time scale versus different cooling timescales.}
    \label{fig:TimeScale}
\end{figure}

Protons that undergo shock acceleration experience energy losses through various mechanisms. In this Appendix, we present a quantitative estimate of  these timescales to show the efficiency of proton acceleration.

The expressions for the acceleration timescale as well as different cooling timescales ($pp$ interaction, adiabatic and diffusion) are provided in Sec.~\ref{sec:model}. In Fig.~\ref{fig:TimeScale}, we have plotted these timescales as a function of proton energy, $E_{\rm p}$ computed at the beginning of CSM interaction, i.e, at $r=r_{\rm i}$.   This figure shows that all the loss timescales become comparable to the acceleration timescale above $10^4$~TeV. In addition to these losses, protons could also lose energy due to photopion production ($p\gamma$), Inverse Compton, Bethe-Heitler, synchrotron radiation. However, the timescales of these losses for the case of SN 2014C are found to be very large ($>10^{13}$ seconds), therefore we do not show them in this figure; see Ref.~\cite{2022JCAP...08..011S} for details. It can be clearly seen from this plot that the relevant loss timescales are long enough for the protons to efficiently accelerate to PeV energies.

\bibliography{biblio}{}
\bibliographystyle{apsrev4-1}

\end{document}